\documentclass[twocolumn,showpacs,preprintnumbers,superscriptaddress,amsmath,amssymb,nofootinbib]{revtex4}
\input psfig.sty
\usepackage{graphicx}

\begin{document}

\title{Dark matter-dark energy interaction for a time-dependent equation of state}

\author{Rafael C. Nunes} 
\affiliation{Departamento de F\'isica, Universidade do Estado do Rio Grande do Norte, 59610-210, Mossor\'o - RN, Brasil}

\affiliation{Funda\c{c}\~ao CAPES, Minist\'erio da Educa\c{c}\~ao, 70040-020, Bras\'{i}lia - DF, Brasil}

\author{Ed\'esio M. Barboza Jr.} \email{edesiobarboza@uern.br}

\affiliation{Departamento de F\'isica, Universidade do Estado do Rio Grande do Norte, 59610-210, Mossor\'o - RN, Brasil}

\date{\today}

\begin{abstract}

In this work we investigate the interaction between dark matter and dark energy for a coupling that obeys the Wang-Meng decaying law, $\rho_{{\rm DM}}\propto (1+z)^{3-\epsilon}$, and the Barboza-Alcaniz dark energy parametric model, $w=w_0+w'_0z(1+z)/(1+z^2)$. Theoretically, we show that   the coupling constant, $\epsilon$, should satisfy the physical constraint $\epsilon\ge0$. We use the most recent data of type Ia supernovae, baryon acoustic oscillations, cosmic microwave background and the Hubble expansion rate function to constrain the free parameters of the model. From a purely observational point of view, we show that is not possible to discard values of the coupling constant in the unphysical region $\epsilon<0$. We show that the uncoupled case, $\epsilon=0$, is in better agreement with the data than any of coupled models in the physical region. We also find that all physically acceptable interaction in dark sector lies in the narrow range $0<\epsilon\le0.034$ ($95\%$ CL). 

\end{abstract}


\maketitle

\section{Introduction}

In the last 15 years there has been a large amount of observational data coming from Type Ia Supernovae (SNe Ia) \cite{acc_exp}, Cosmic Microwave Radiation Background (CMB) \cite{cmb} and Large Scale Structure (LSS) \cite{lss} shown that the expansion rate of the Universe is increasing. Finding out the causes of this acceleration has been the biggest challenge to cosmologists. In order to keep general relativity untouched, a fluid of negative pressure, dubbed dark energy (DE), must be added to the universe content to yield an acceleration. In this scenario, the cosmological constant proposed by Einstein emerges as the most appealing candidate to DE since it acts on the field equations like a fluid with $p_{\Lambda}=-\rho_{\Lambda}$ and can be associated with the zero point energy of the quantum fields. However, in spite of its agreement with the majority of cosmological data, the cosmological constant leads to a tremendous discrepancy between theory and observation: its observed value is at least $60$ orders of magnitude lower than the theoretical value provided by the quantum field theory \cite {weinberg}.  This enormous discrepancy has made DE models beyond the cosmological constant widely studied. Such models presume that some unknown symmetry cancels out the vacuum energy contribution. If the vacuum energy cannot be canceled, another attempt to alleviate the conflict between theory and observation is to assume that the cosmological constant evolves with time. Such an assumption means that dark matter (DM) and the vacuum energy are not conserved separately. Due the success of the cosmological term in explaining the current observations, phenomenological dark energy models, frequently characterized by the ratio between pressure and density, $w\equiv p_{{\rm DE}}/\rho_{{\rm DE}}$, and vacuum decay scenarios are almost always built to get the standard $\Lambda$CDM model as an special case. A most general approach can be achieved by assuming an interaction between DE and DM. 

In this paper we study an interaction scenario where the DE is described by the equation of state (EoS) parameter \cite{barboza}
\begin{equation}
\label{EoS}
w(z)=w_0+w'_0\frac{z(1+z)}{1+z^2}
\end{equation}
and the DM density follows the Wang-Meng evolution law \cite{wang-meng}:
\begin{equation}
\label{wang-meng}
\rho_{{\rm DM}}=\rho_{{\rm DM},0}(1+z)^{3-\epsilon}.
\end{equation}

\noindent In the above equations the subscript $0$ denotes the current value of a quantity, the prime denotes differentiation with respect to the redshift $z$ and $\epsilon$ is a constant that quantifies the matter dilution due the interaction.

The main advantage of the EoS parameterization (\ref{EoS}) is that it is a well behaved function of the redshift during the entire history of the universe ($z\in[-1,\infty[$) which allows one to enclose in its functional form the important case of a quintessence scalar field ($-1<w(z)<1$) \cite{quintessence}.  By noting that $w(z)$ has absolute extremes in $z_{\pm}=1\pm\sqrt{2}$ corresponding, respectively, to $w_-=w(z_-)=w_0-0.21w'_0$ and $w_+=w(z_+)=w_0+1.21w'_0$, it is possible to divide the parameter space $(w_0,w'_0)$ into defined regions associated with distinct dark energy models which can be confronted with the observational constraints to confirm or rule out a given DE model. For $w'_0>0$, $w_-$ is a minimum and $w_+$ is a maximum and for $w'_0<0$ this is inverted. Since for quintessence and phantom \cite{phantom} scalar fields the EoS is limited by $-1\leq w(z)\leq1$ and $w(z)<-1$, respectively, the region occupied in the $(w_0,w'_0)$ plane by these fields can be determined easily. For quintessence we get $-1\leq w_0-0.21w'_0$ and $w_0+1.21w'_0\leq1$ if $w'_0>0$ and $-1\leq w_0+1.21w'_0$ and $w_0-0.21w'_0\leq1$ if $w'_0<0$. For phantom fields we get $w'_0<-(1+w_0)/1.21$ if $w'_0>0$ and $w'_0>(1+w_0)/0.21$ if $w'_0<0$. Points out of these bounds corresponds to DE models that have crossed or will cross the phantom divide line.

This {\it paper} is organized as follows: in Section II the basic equations employed in the analysis are developed; in Section III the constraints on the parameters $w_0$, $w'_0$ and $\varepsilon$ are obtained observationally from current SNe Ia, BAO, $H(z)$ and CMB data; in Section IV we obtain the quintessence and phantom scalar field description for the model under consideration and in Section V we present our conclusions and final comments. 

\begin{figure*}[t]
\centerline{\psfig{figure=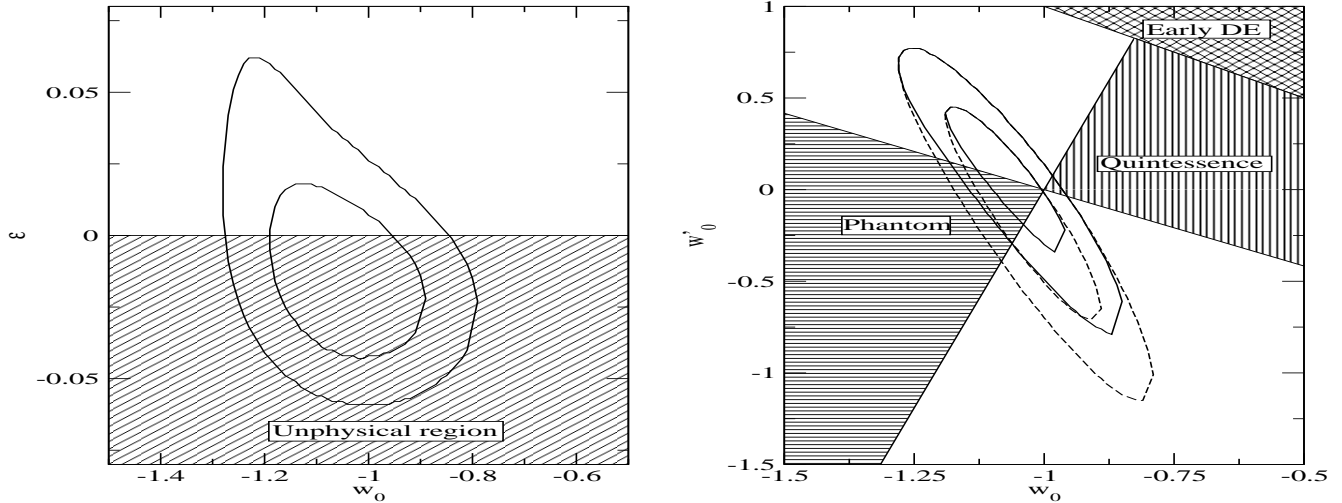,width=7.4truein,height=3.0truein,angle=270}
\hskip 0.1in}
\caption{The $w_0-\epsilon$ (left) and $w_0-w'_0$ (right) parametric spaces. The blank regions in the $w_0-w'_0$ plane indicate models that at some point of the cosmic evolution have switched or will switch from quintessence to phantom behaviors or vice-versa. The Early DE region corresponds to the region where DE dominates over matter in early times. The dashed contours in the $w_0-w'_0$ plane are the ones obtained when we allow that $\epsilon<0$. The contours are drawn for $\Delta \chi^2 = 2.30$ and 6.17.\label{fig:1}}
\end{figure*}

\section{The interaction model}

By assuming that DM and DE are not conserved separately, stress-energy conservation $\nabla_{\nu}T^{\mu\nu}=0$ gives

\begin{equation}
\label{conservation}
\dot{\rho}_{{\rm DM}}+3\frac{\dot{a}}{a}\rho_{{\rm DM}}=-\dot{\rho}_{{\rm DE}}-3\frac{\dot{a}}{a}(1+w)\rho_{{\rm DE}},
\end{equation}

\noindent where $a$ is the scale factor and the dot denotes time derivatives. For a dark matter density evolving according (\ref{wang-meng}) we get, from the above equation, that the dark energy density evolves as

\begin{equation}
\label{de_density}
\rho_{{\rm DE}}(z)=f_u(z)\Big[\rho_{{\rm DE},0}+\epsilon\rho_{{\rm DM},0}\int_0^z dz\frac{(1+z)^{2-\epsilon}}{f_u(z)}\Big],
\end{equation}
where 
\begin{equation}
\label{uncoupled0_f}
f_u(z)=(1+z)^{3}\exp\Big[\int_0^z\,dz\frac{w(z)}{1+z}\Big]
\end{equation}
is the ratio between $\rho_{{\rm DE}}$ and $\rho_{{\rm DE},0}$ for the uncoupled case. For the EoS (\ref{EoS}), we get
\begin{equation}
\label{uncoupled_f}
f_u(z)=(1+z)^{3(1+w_0)}(1+z^2)^{3w'_0/2}.
\end{equation}
If $\epsilon<0$, the second term of eq. (\ref{de_density}) becomes negative and may dominate over the first making $\rho_{{\rm DE}}(z)<0$, leading to an unphysical situation. In order to make this point clearer, let us examine the asymptotic limit. For large redshifts, $f_u\to z^{3(1+w_0+w'_0)}$ and
\begin{equation}
\label{assmp_density}
\rho_{{\rm DE}}(z)\to\rho_{{\rm DE},0}z^{3(1-\vert w_0+w'_0\vert)}+\frac{\epsilon\rho_{{\rm DM},0}z^{3-\epsilon}}{3\vert w_0+w'_0\vert-\epsilon}.
\end{equation}
Thus, if $\epsilon<0$, the first term varies with a power lower than $3$ while the second term varies with a power higher than $3$ for large redshifts and the sign of $\epsilon$ will define the sign of $\rho_{{\rm DE}}$. Since a negative density corresponds to an unphysical solution, the case $\epsilon<0$ can be discarded. The same constraint was obtained in \cite{JJ} from thermodynamics arguments for the vacuum decay case. Note that this result is valid for all classes of parametric DE models where the conditions $w(z\gg1)\to constant$ and $f_u(z\gg1)\to z^{constant}$ are satisfied as, for instance, the Chevalier-Polarski-Linder (CPL) model. 

For the model under consideration, the Friedmann' equation becomes
\begin{eqnarray}
\label{friedmann_eq}
H^2&=&H^2_0\big[\Omega_{\gamma,0}(1+z)^4+\Omega_{{\rm b},0}(1+z)^3+\nonumber\\
      &+&\Omega_{{\rm DM},0}(1+z)^{3-\varepsilon}+\Omega_{\kappa,0}(1+z)^2+\nonumber\\
      &+&(1-\Omega_{\gamma,0}-\Omega_{{\rm b},0}-\Omega_{{\rm DM},0}-\Omega_{\kappa,0})f(z)\big],
\end{eqnarray}
where $H\equiv\dot{a}/a$ is the Hubble parameter, $f(z)\equiv\rho_{{\rm DE}}/\rho_{{\rm DE},0}$, $\Omega_{\gamma,0}=\rho_{\gamma,0}/\rho_{c,0}$, $\Omega_{{\rm b},0}=\rho_{{\rm b},0}/\rho_{c,0}$ and $\Omega_{{\rm DM},0}=\rho_{{\rm DM},0}/\rho_{c,0}$ are, respectively, the density parameters of radiation, baryonic matter and dark matter with $\rho_{c,0}=3H_0^2c^2/8\pi G$ and $\Omega_{\kappa,0}=-\kappa c^2/(a_0H_0)^2$ is the curvature parameter. Motivated by the recent results of the CMB power spectrum \cite{planck}, we assume spatial flatness in the following analyses.

\section{Observational Constraints}

In order to discuss the current observational constraints on $w_0$,  $w'_0$ and $\epsilon$, the Union 2.1 SN Ia sample of  Ref.~\cite{union}, which is an update of the Union 2 compilation and comprises 580 data points \cite{union}, is used. Along with the SNe Ia data, and to diminish the degeneracy between the parameters $w_0$,  $w'_0$ and $\epsilon$, we use $28$ measurements of the Hubble function $H(z)$ \cite {hdata}, and results of BAO and CMB experiments. For the BAO measurements, the six estimates of the BAO parameter 
\begin{equation}
\label{BAO_parameter}
{\cal{A}} (z) = D_V{\sqrt{\Omega_{\rm{m},0} H_0^2}}
\end{equation}
given in Table 3 of Ref.~\cite{blakea} are used. In this latter expression, 
$D_V = [r^2(z_{\rm{BAO}}){z_{\rm{BAO}}}/{H(z_{\rm{BAO}})}]^{1/3}$
is the so-called dilation scale, defined  in terms of the dimensionless comoving distance $r$. For the CMB, only the
measurement of the CMB shift parameter~\cite{wmap}
\begin{equation}
\label{shift_parameter}
{\cal{R}} = \Omega_{\rm{m},0}^{1/2}r(z_{\rm{CMB}}) = 1.725 \pm 0.018\;,
\end{equation}
where $z_{\rm{CMB}} = 1089$ is used. In both (\ref{BAO_parameter}) and (\ref{shift_parameter}), $\Omega_{\rm{m},0}=\Omega_{\rm{b},0}+\Omega_{\rm{DM},0}$. Thus, in the present analyses, the function  $\chi^2 = \chi^{2}_{\rm{SNe}} + \chi^{2}_{\rm{H}} + \chi^{2}_{\rm{BAO}} + \chi^{2}_{\rm{CMB}},$ 
which takes into account all the data sets mentioned above, is minimized. The present value of the Hubble parameter $H_0$ is marginalized and the dark matter density parameter $\Omega_{\rm{DM},0}$ is kept fixed at $0.24$ in our analysis.

Figure \ref{fig:1} shows the results of the statistical analysis within $68\%$ and $95\%$ confidence levels. The left figure 
shows the $w_0-\epsilon$ parameter space obtained by marginalizing over $w'_0$ and the right figure shows the $w_0-w'_0$ parameter space obtained marginalized over $\epsilon$. We leave $\epsilon$ free to run to any value. The best fit values are $w_0=-1.04^{+0.09}_{-0.10}$, $w'_0=-0.11^{+0.38}_{-0.38}$ and $\epsilon=-0.016^{+0.021}_{-0.017}$ with the upper and lower values denoting the one parameter $1\sigma$ errors. The solid (dashed) lines in the $w_0-w'_0$ space corresponds to the contours obtained for $\epsilon\ge0$ ($\epsilon$ free). As we can see, only a small portion of the $w_0-w'_0$ confidence regions lies in the quintessence region with the largest portion occupied by models the have crossed or will cross the phantom divide line. Also, the largest portion of the $\epsilon$ values allowed by the data lies in the unphysical region. We can interpret this result as lack of sensitivity of the data to the physical constraint $\epsilon\ge0$. In fact, the uncoupled case is favored over the coupled case. In order to make this point clearer, we list in Table \ref{tab:1} some values of the parameters $w_0$ and $w'_0$ obtained for some values of $\epsilon\ne0$ and compare its $\chi^2_{min}$ with the $\chi^2_{min}$ of the uncoupled case ($\epsilon=0$). Values of $\Delta\chi^2=\chi^2_{{\rm coupled}}-\chi^2_{{\rm uncoupled}}<0$ means that the coupled case provides a better fit to data than the uncoupled case. As can be seen, only in the narrow range $-0.03<\epsilon<0$ inside the unphysical region is the coupled case in better agreement with the data than the uncoupled scenario. Outside this small interval, the uncoupled case is favored by the data. If we impose the constraint $\epsilon\ge0$, we find, for one parameter, that $0\le\epsilon\le0.034$ in $2\sigma$. This is a very small range of the physical region where a coupling making sense. This would be a strong argument to discard a coupling producing the dark matter density law (\ref{wang-meng})\footnote{Remember that from WMAP tree years \cite{cmb} constraint on the curvature parameter, $-0.023\le\Omega_{k,0}\le0.001$, the majority of physicists began to adopt $\Omega_{k,0}=0$.}. However, due the lack of sensitivity of the data to the physical constraint $\epsilon\ge0$ and the fact that we are concerned only to a DE model whose the EoS parameter satisfies (\ref{EoS}), we cannot make such a strong statement and, therefore, the possibility of a coupling in dark sector producing a dilution of dark matter according (\ref{wang-meng}) remains open.

\begin{figure*}[t]
\centerline{\psfig{figure=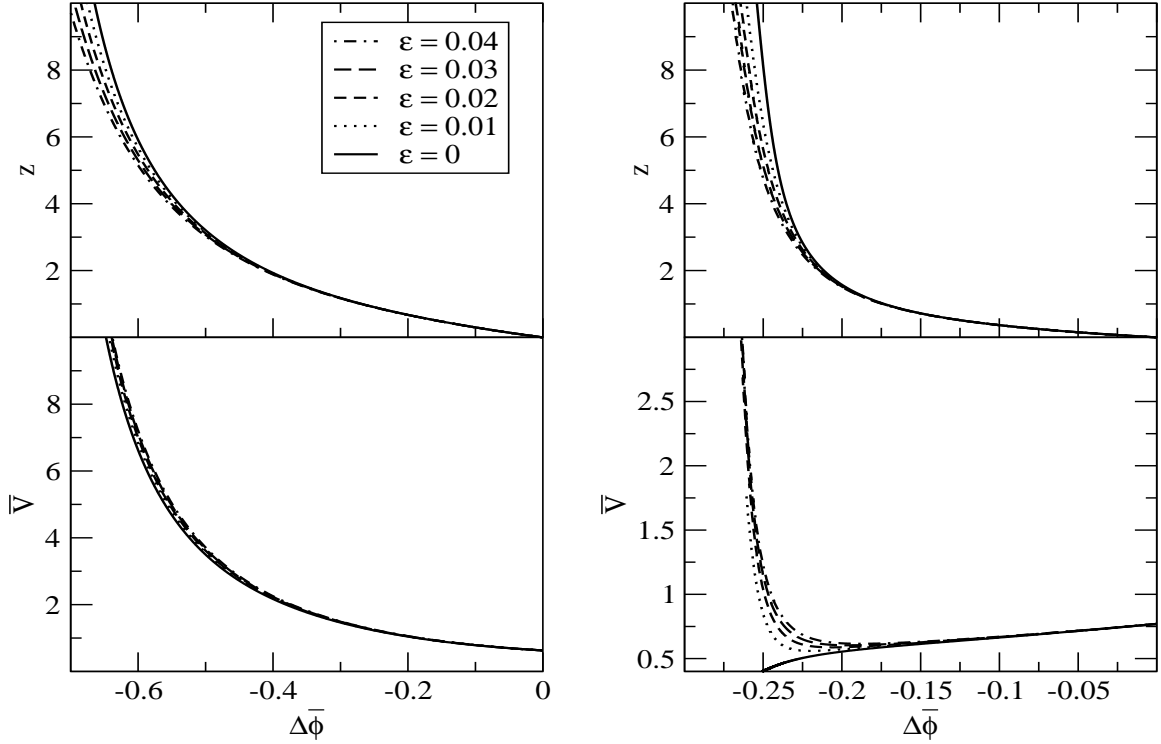,width=7.4truein,height=5.0truein,angle=270}
\hskip 0.1in}
\caption{Scalar field description of the coupled case for five selected points in the quintessence (left) and phantom (right) regions. The uncoupled case also is shown (full lines). The values of $(w_0,w'_0)$ are $(-0.8, 0.2)$ for quintessence and $(-1.2, 0.1)$ for phantom fields.\label{fig:2}}
\end{figure*}

\begin{table}[!h]
	\begin{center}
		\begin{tabular}{ccccc}
		\hline
		\hline\\
		$\epsilon$&$ w_0$&$w'_0$& $ \chi^2_{min}$&$\Delta\chi^2_{min}$\\
		\hline
		\hline\\
                   -0.04&$-1.01^{+0.14}_{-0.14}$&$-0.32^{+0.50}_{-0.56}$&$583.94$&$1.21 $\\
                   -0.03&$-1.02^{+0.14}_{-0.14}$&$-0.24^{+0.48}_{-0.56} $&$582.75 $&$0.02 $\\
		-0.02&$-1.04^{+0.15}_{-0.13}$&$-0.13^{+0.45}_{-0.57}$&$582.20 $&$-0.53 $\\
		-0.01&$-1.05^{+0.14}_{-0.13}$&$-0.05^{+0.45}_{-0.53}$&$582.22 $&$ -0.51$\\
                   0.00&$-1.08^{+0.15}_{-0.12}$&$0.10^{+0.39}_{-0.55}$&$582.73 $&$ 0.00$\\
                   0.01&$-1.10^{+0.14}_{-0.12}$&$0.22^{+0.36}_{-0.50}$&$583.58$&$ 0.85$\\
		\hline
		\hline
		\end{tabular}
	\end{center}
	\caption{Comparative analysis between coupled ($\epsilon\ne0$) and uncoupled case ($\epsilon=0$). Values of $\Delta\chi^2=\chi^2_{{\rm coupled}}-\chi^2_{{\rm uncoupled}}<0$ means that the coupled case provides a better fit to data than uncoupled case. The coupled case provides a better fit than uncoupled case only in the range $-0.03<\epsilon<0$ which is in the unphysical region.}
	\label{tab:1}
\end{table}

\section{Scalar field description}

As we have already stressed, there are regions of the parametric space where the EoS parameter is associated with  a quintessence scalar field ($w(z)\in[-1,1]\,\forall z$) and regions where the EoS parameter is associated with  a phantom field ($w(z)<-1\,\forall z$). Thus, for completeness, we construct the DE potential $V(\phi)$ directly from the EoS for both quintessence and phantom cases.

Now, the energy density and pressure of the dark energy field are given by
\begin{subequations}
\begin{equation} \label{rhophi}
\rho_{{\rm DE}} = \alpha\frac{1}{2}\phi^2 + V(\phi)\;,
\end{equation}
\begin{equation} \label{pphi}
p_{{\rm DE}} = \alpha\frac{1}{2}\phi^2 - V(\phi)\;,
\end{equation}
\end{subequations}
where $\alpha=\pm 1$ stands for quintessence ($-1 < w_{{\rm DE}} \leq -1/3$) and phantom fields ($w_{{\rm DE}} < -1$), respectively.

By combining Eqs. (\ref{rhophi}) and (\ref{pphi}), we obtain
\begin{subequations}
\begin{equation}
\label{dphi}
\dot{\phi}^2=\frac{1+w_{{\rm DE}}}{\alpha}\rho_{{\rm DE}}\;,
\end{equation}
and
\begin{equation}
\label{vphi}
V(\phi)=\frac{1}{2}(1-w_{{\rm DE}})\rho_{{\rm DE}}\;,
\end{equation}
or, in terms of $z$,
\begin{equation}
\dot{\phi}=\frac{d\phi}{dz}\dot{z}=-\frac{d\phi}{dz}(1+z)H(z)\;,
\end{equation}
so that,
\begin{equation}
\frac{d\phi}{dz}=\pm\frac{1}{(1+z)H(z)}\sqrt{\frac{1+w_{{\rm DE}}}{\alpha}\rho_{{\rm DE}}}\;,
\end{equation}
\end{subequations}
where the negative (positive) signs stands to $\dot{\phi}>0$ ($\dot{\phi}<0$). Here, we adopt the negative sign.

By defining $\bar{\phi}\equiv\sqrt{8\pi\,G/3}\phi$ and $\bar{V}\equiv V/\rho_{c,0}$ and taking into account that $(1+w_{{\rm DE}})/\alpha=\vert1+w_{{\rm DE}}\vert$, we have
\begin{subequations}
\begin{eqnarray}
\label{phitilde}
\Delta\bar{\phi}&\equiv&\bar{\phi}-\bar{\phi}_0 \nonumber\\ &=&-\int_0^z\frac{1}{(1+z)\eta(z)}\sqrt{\vert1+w_{{\rm DE}}(z)\vert\Omega_{{\rm DE,0}}f(z)}\nonumber\\&&
\end{eqnarray}
and
\begin{equation}
\label{vphitilde}
\bar{V}(\bar{\phi})=\frac{1}{2}[1-w_{{\rm DE}}(z)]\Omega_{{\rm DE,0}}f(z),
\end{equation}
\end{subequations}
where $\eta(z)=H(z)/H_0$ and $f(z) = \rho_{{\rm DE}}/\rho_{{\rm DE,0}}$ is given by (\ref{de_density}). Note also that Eqs. (\ref{phitilde}) and (\ref{vphitilde}) are valid for both quintessence and phantom fields.

By combining numerically Eqs. (\ref{EoS}), (\ref{friedmann_eq}), (\ref{phitilde}) and (\ref{vphitilde}), and taking into account the above constraints, we obtain the scalar field $\Delta{\bar\phi}$ and the resulting potential ${\bar V} ({\bar\phi})$ for quintessence and phantom cases. Figure \ref{fig:2}
shows the evolution of the dark energy field as function of the redshift (top panels) and the potential as function of the field (bottom panels)  for quintessence (left) and phantom (right) regimes for some selected values of $\epsilon$. For simplicity, we consider two pairs of values $(w_0,w'_0)$, that are $(-0.8, 0.2)$ and $(-1.2, 0.1)$, corresponding to quintessence and phantom behaviors, respectively. We also plot the scalar field potential for the uncoupled case (full lines). As we can see, when $\epsilon$ decreases, the quintessence scalar field rolls more smoothly until the minimum of its potential. In the phantom regime we note that the potential evolves in a different manner for the coupled and uncoupled cases. For the coupled case ($\epsilon>0$) the potential increases with $z$ while for uncoupled case ($\epsilon=0$) the potential decreases with $z$ and goes to zero when $z\to\infty$. This can be explained as follows: for large redshift values, the potential evolves as

\begin{eqnarray}
\label{assymtoptic}
\bar{V}(z)&\to& \frac{1}{2}(1+\vert w_0+w'_0\vert)\Big[\Omega_{{\rm DE,0}}z^{3(1-\vert w_0+w'_0\vert)}\nonumber\\
&+&\frac{\epsilon\Omega_{{\rm DM,0}}z^{3-\epsilon}}{3\vert w_0+w'_0\vert-\epsilon}\Big]
\end{eqnarray}

\noindent so, for phantom fields ($\vert w_0+w'_0\vert>1$) the first term goes to zero for large $z$ while the second term becomes large and dominates the functional form of the potential.

\section{Final Remarks}

We have examined theoretical and observational aspects of models in which dark energy interacts with dark matter. We have assumed that the dark energy is described by a time dependent EoS parameterization given by Eq. (\ref{EoS})~\cite{barboza} and that the dark matter evolves according the Wang-Meng law (\ref{wang-meng}) \cite{wang-meng}. Theoretically, we have derived that the coupling constant should be in the range $\epsilon\ge0$. We have performed a joint statistical analysis involving some of the most recent cosmological measurements of SNe Ia, BAO peak, CMB shift parameter and $H(z)$. From a purely observational perspective, we have shown that the hypothesis of a coupling in the dark sector cannot be ruled out. However,  we have shown that, for the model investigated, only in the narrow range $-0.03\le\epsilon<0$ inside the unphysical region the interaction scenario is in a better agreement  with the data than the uncoupled case. In $2\sigma$, physical values of the coupling constant are in the range $0<\epsilon\le0.034$. We have also noted that the observational data employed in our analysis shows no sensitivity to the physical constraint $\epsilon\ge0$. 

Also, following the recipe given in ~\cite{method,ebej}, we have derived the scalar field description for this $w(z)$ parameterization for quintessence and phantom fields. We have shown that the main difference between the uncoupled and coupled cases occurs in the phantom regime, while in the first case the scalar field potential decrease with $z$, whereas in the second case the opposite situation occurs.

\begin{acknowledgments} 
RCN acknowledges financial support from CAPES Scholarship Bex 13222/13-9. The authors are very grateful to Thomas Dumelow and Ja\'{i}lson  Alcaniz for a critical reading of the manuscript and useful comments. 
\end{acknowledgments}

\end{document}